\pdfoutput=1
\RequirePackage{fix-cm}
\documentclass[reqno]{amsproc}
\usepackage{amssymb}
\usepackage{hyperref}
\usepackage{microtype}

\usepackage{graphicx}

\usepackage[citation-order,nobysame]{amsrefs}
\usepackage{xyzbib}

\usepackage{mathtools}
\mathtoolsset{showonlyrefs}

\numberwithin{equation}{section}
\let\cite=\cites
\newcommand{\rmd}{\mathrm{d}}
\newcommand{\rmi}{\mathrm{i}}
\newcommand{\rme}{\mathrm{e}}

\newcommand{\thetac}{\theta_\mathrm{c}}
\newcommand{\omegac}{\omega_\mathrm{c}}

\renewcommand{\figurename}{Fig.}

\DeclareMathOperator*{\res}{\mathrm{res}}

\begin{document}


\title{Scaling limit of domino tilings on a pentagonal domain}

\author{Filippo Colomo}
\address{INFN, Sezione di Firenze, 
Via G. Sansone 1, I-50019 Sesto Fiorentino (FI), Italy}
\email{colomo@fi.infn.it}

\author{Andrei G. Pronko}
\address{Steklov Mathematical Institute, 
Fontanka 27, 191023 St.\,Petersburg, Russia}
\email{a.g.pronko@gmail.com}

\begin{abstract}
We consider the six-vertex model at its free-fermion point with domain
wall boundary conditions, which is equivalent to random domino tilings
of the Aztec diamond. We compute the scaling limit of a particular
non-local correlation function, essentially equivalent to the
partition function for the domino tilings of a pentagon-shaped domain,
obtained by cutting away a triangular region from a corner of the
initial Aztec diamond. We observe a third-order phase transition when
the geometric parameters of the obtained pentagonal domain are tuned
to have the fifth side exactly tangent to the arctic ellipse of the
corresponding initial model.
\end{abstract}


\maketitle

\section{Introduction}

Random tilings of regular lattices and related dimer models possess
numerous fascinating features, among those, perhaps, most intriguing
and important are phase separation phenomena. A seminal example is
provided by the tilings of an hexagon by lozenges (rhombi), which are
in bijection with boxed plane partitions \cite{CLP-98}.  Another
classic instance is given by the random tilings of the `Aztec diamond'
by dominoes, where one can observe ordered, or `frozen',
configurations in the four corners, outside a central disordered
region. The phase separating curve resulting in the scaling limit is
named arctic circle, or, for a weighted counting of configurations,
arctic ellipse \cite{JPS-98,CJY-12}. In rather general settings these
problems can be studied in the framework of dimer
models \cite{KOS-06,KO-06}.

Phase separation phenomena may be also observed in vertex models,
which can be viewed as interacting generalizations of dimer models
\cite{EKLP-92a,EKLP-92b}. In particular, when the parameters
(Boltzmann weights) of the vertex models are tuned to the
non-interacting case, these can be used as an alternative tool to
study random tilings. Consequently, many results obtained in the
context of the vertex models find application in various problems in
combinatorics and related areas, including the theory of symmetric
polynomials, see, e.g., \cite{MPP-22,ABW-23} and references therein.

As an example, one can consider the domino tilings of the Aztec
diamond with a cut-off corner of macroscopic square shape and given
size, and study the bulk properties of the tilings as the size is
varied \cite{CP-13}.  In this problem one can rely on the well-known
correspondence between the domino tilings of the Aztec diamond and the
six-vertex model with domain wall boundary conditions
(DWBC) \cite{K-82,EKLP-92a,EKLP-92b}. In \cite{CP-13}, this
correspondence have been applied to study thermodynamics of the domino
tilings of the Aztec diamond with a cut-off corner of a square shape.
The six-vertex model have been considered with a square portion of the
lattice removed, again with DWBC, and the resulting domain has the
L-letter shape \cite{CP-15}.  The Boltzmann weights have to be chosen
to obey the free-fermion condition.

In the present paper, we apply similar ideas to study thermodynamics
of the domino tilings of the Aztec diamond with a cut-off corner of a
triangular shape, the resulting domain having a pentagonal shape.  Our
main result is the expression for the free energy of the random domino
tilings of the resulting region.  It is worth mentioning that domino
tilings of a pentagonal domain has already been considered
in \cite{FV-21}, see also \cite{FV-17}. However the focus there was
rather on a discrete microscopic derivation of the hard-edge tacnode
process \cite{DV-15}. 

We use the observation \cite{CP-07b} that the partition function of
the six-vertex model on a modified lattice can be written as certain
non-local correlation function of the six-vertex model on the original
lattice with DWBC.  To obtain a cut-off corner of rectangular shape
one has to consider the correlation function known as the emptiness
formation probability (EFP), which can be viewed as a test function
for total ferroelectric order in a rectangular sub-region in a corner
of the original lattice. To extend these ideas to a cut-off corner of
arbitrary shape, one can introduce the generalized emptiness formation
probability (GEFP) \cite{CPS-16}. Under certain choice of the
geometric parameters, it can produce a cut-off region of generic
shape, for example, a triangular one. Here, we focus on this case and
derive the behavior of GEFP in the thermodynamic limit, for the
free-fermion Boltzmann weights.

We use the method of log-gas to derive the free energy, or,
equivalently, the leading term of the logarithm of the GEFP, following
the techniques exposed in detail in \cite{CP-13,CP-15}.  It has to be
noted that in the case of EFP, related to domino tilings of the
L-shaped domain, there exists a connection to the theory of the sixth
Painlev\'e equation, which makes it possible to build thermodynamic
limit asymptotic expansions up to an arbitrary order in the large
parameter \cite{KP-16}; see also \cite{BP-24}, where an application of
this method to the five-vertex model with scalar-product boundary
conditions was considered.  For the pentagonal domain such a
connection is unknown.  Nevertheless, a random-matrix-like formula is
available for the partition function \cite{KP-19}, which, despite
some relevant difference, may actually be treated with random matrix
model technologies.  We use this formula as the starting point in our
considerations here.

Restating the result in terms of tilings, we get the free energy
density for the domino tilings of the Aztec diamond with a triangular
cut-off corner, as a function of the size of the cut-off triangle. We
observe that the free energy displays a third-order phase transition
of Douglas--Kazakov type \cite{DK-93}, when the geometric parameters
of the obtained pentagonal domain are tuned to have the fifth side
exactly tangent to the arctic ellipse, that is the phase separation
curve of the original (unmodified) Aztec diamond.  In other words,
phase separation curves can be viewed as critical curves in the space
of parameters describing the macroscopic geometry of the tiled
region \cite{CP-13,CP-15}.

\section{The generalized emptiness formation probability}

\begin{figure}
\centering
\includegraphics{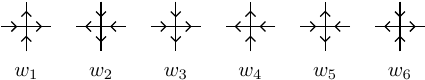}
\caption{Arrow configurations around a vertex (first row) 
and the corresponding vertex Boltzmann weights (second row).}
\label{fig-SixVertices}
\end{figure}

We consider the six-vertex model on the $N\times N$ lattice.  This
means that the lattice is obtained from the intersection of $N$
horizontal and $N$ vertical lines. The configurations of the model are
obtained by orienting the edges (placing arrows on them), with the
condition that around each vertex two arrows point inward and two
outward. This condition is known as the ice rule, and allows for
exactly six possible vertex configurations, to which we assign the
Boltzmann weights $w_j$, $j=1,\dots,6$, as in
\figurename~\ref{fig-SixVertices}.

We consider the model with DWBC, that is, we require the arrows to
point outward on the horizontal external edges, and inward on the
vertical ones \cite{K-82,I-87,ICK-92}.  We label the vertex at the
intersection of the $r$th vertical line (counted from the right) and
the $s$ horizontal line (counted from the top) by the lattice
coordinates $(r,s)$, with $r,s=1,\dots,N$.  The partition function is
defined as usual as the sum over all possible configurations of arrows
on the lattice, satisfying the ice-rule and the boundary conditions.

To define the GEFP, let us consider the set of $s$ horizontal edges
$\{e_1,\dots,e_s\}$, where $s\leq N$, with the edge $e_j$ lying to
left of the vertex of coordinates $(r_j,j)$, $j=1,\dots,s$. The
$r_j$'s are required to be a weakly increasing sequence, $1\leq
r_1\leq\dots\leq r_s\leq N$.  Note that one may associate to the above
set of edges an $s$-row Ferrer diagram, defined by the partition
$\lambda_s=(N-r_1,\dots,N-r_s)$.

\begin{figure}
\centering
\includegraphics{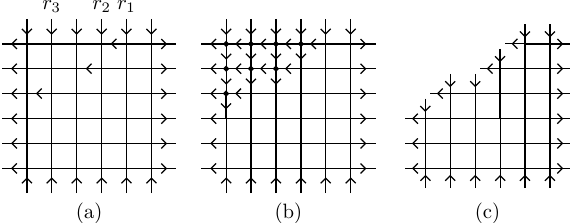}
\caption{GEFP for the $N\times N$ lattice: (a) The configuration of arrows,
whose probability is described by GEFP; (b)  
The Ferrer diagram $\lambda_s$ associated with the 
``frozen'' region with dots standing for vertices of type 2; 
(c) The cut-off domain obtained by removing the vertices forming $\lambda_s$. 
Here $N=6$, $s=3$, $(r_1,r_2,r_3)=(2,3,5)$, and
$\lambda_s=(4,3,1)$.}
\label{fig-GEFPbis}
\end{figure}

The GEFP, denoted $G_{N,s}^{(r_1,\dots,r_s)}$, is defined as the
probability of having the arrows on the $s$ horizontal edges
$e_1,\dots,e_s$ all pointing leftward,
see \figurename~\ref{fig-GEFPbis}a.  The GEFP may equivalently be
defined as the probability of observing in the top-left corner of the
$N\times N$ lattice a frozen region of shape $\lambda_s$, with all
vertices therein being of type 2, see \figurename~\ref{fig-SixVertices}b.
The equivalence of the two definitions follows from both the ice-rule
and the DWBC. Note that, the top-left corner being frozen, it may as
well be cut away, allowing to relate GEFP directly to the partition
function of the six-vertex model on a quite general class of domain on
the square lattice, still with domain wall boundary conditions,
see \figurename~\ref{fig-GEFPbis}c.

It follows from the definition of GEFP that it may be evaluated as the
ratio between two different partition functions of the model: one with
the boundary conditions implementing the freezing of the assigned
Ferrer diagram, and one with DWBC. In \cite{CPS-16}, a multiple
integral representation has been derived for GEFP, by computing these
two partition functions in the framework of the quantum inverse
scattering method \cite{KBI-93}. 

When the $r_j$'s are chosen to be all equal to $r$, the GEFP reduces
to the EFP, $F_N^{(r,s)}\equiv G_{N,s}^{(r,\dots,r)}$, giving
the probability of observing a frozen rectangle of size $(N-r)\times
s$ in the top-left corner of the lattice.  Clearly in this case some
simplifications occur, allowing for the derivation of various multiple
integral representations \cite{CP-07b,CP-12,P-13,CDP-21}, and for an
analytic treatment of their behaviour in the scaling
limit \cite{CP-09}.  If one further specialize $r=N-s$, the frozen
region becomes an $s\times s$ square, with additional simplifications
allowing for the derivation of some more explicit
results \cite{CP-24}.

Another interesting situation, allowing for some simplification as
well, is that in which the region required to be frozen, and the
corresponding Ferrer diagram, have a triangular shape. This
is realized by specializing $r_j=N-s+j-1$, $j=1,\dots, s$, with
$s<N$. We may thus define the Triangular Domain EFP (TDEFP),
\begin{equation}
T_{r,s}:=G_{N,s}^{(N-s,N-s+1,\dots,N-1)}, \qquad r+s=N,
\end{equation}
giving the probability of observing in the top-left corner of the
$N\times N$ lattice, a frozen triangular region of size $s$
\cite{KP-19}.

The problem we want to address is to investigate TDEFP in the scaling
limit, for the six-vertex model at its
free-fermion point.  As a by-product, we  derive the free energy per site of the
six-vertex model on a pentagonal domain, or, equivalently, the
free energy for domino tilings of the pentagonal domain, 
that we will exposed in detail below.

\section{The free-fermion  case}

In what follows we
restrict ourselves to the case where the Boltzmann
weights obey the free-fermion condition:
\begin{equation}\label{FF}
w_1w_2+w_3w_4=w_5w_6.
\end{equation}
Under this condition, the partition function of the six-vertex model
on the $N\times N$ lattice with DWBC, $Z_N$, takes a very simple form,
see, e.g., \cite{EKLP-92a,EKLP-92b}:
\begin{equation}\label{Z_FF}
Z_N=w_5^{\frac{N(N-1)}{2}}w_6^{\frac{N(N+1)}{2}}.
\end{equation}
We parameterize the Boltzman weights as follows:
\begin{equation}\label{weightsFF}
w_1=w_2=\sqrt{\rho(1-\alpha)},\qquad
w_3=w_4=\sqrt{\rho\alpha},\qquad
w_5=1,\qquad
w_6=\rho.
\end{equation}
Recall that, due to the DWBC, in any given configuration of the model
the number of vertices of type 6 minus that of type 5 equals
$N$. Consequently, the parameter $\rho$ is just an overall
normalization of weights. Instead, the parameter $\alpha\in[0,1]$ is
relevant, favoring the occurrence of vertices of type 1 and 2
($\alpha<1/2$), or of type 3 and 4 ($\alpha>1/2$).

Under the free-fermion condition \eqref{weightsFF}, the multiple
integral representation for the GEFP simplifies significantly, and may
be rewritten as a determinant, see
\cite{KP-19}, Eq.~(6). One has:
\begin{equation}\label{detGEFPalpha}
G_{N,s}^{(r_1,\dots,r_s)}=\left(\prod_{j=1}^{s}(1-\alpha)^{N-r_j} \right)
\det_{1\leq j,k\leq s}\left[P_{N-s+j+k-2}^{(N-r_{s-j+1},j)}(\alpha)\right]
\end{equation}
with
\begin{equation}
P_l^{(m,n)}(\alpha):=\oint\frac{x^l}{(x-\alpha)^m(x-1)^n}\frac{\rmd x}{2\pi\rmi},
\end{equation}
where the integration is taken over a positive contour enclosing the
poles at $\alpha$ and 1. It may be verified that $P_l^{(m,n)}(\alpha)$
is a polynomial of degree $l-m-n+1$ in $\alpha$.

In the case of TDEFP, \eqref{detGEFPalpha} simplifies further, and
various alternative representations have been proposed in
\cite{KP-19}.  Below we shall resort to two of them.
The first one follows directly from the determinantal representation
\eqref{detGEFPalpha} by specializing $r_j=N-s+j-1$,
\begin{equation}\label{TDEFP_det}
  T_{r,s}=(1-\alpha)^{s(s+1)/2}\det_{1\leq j,k\leq s}
  \left[P_{r+j+k-2}^{(j,j)}(\alpha)\right], 
\end{equation}
and we recall that $r\equiv r_1= N-s$.

The second representation reads:
\begin{equation}\label{TDEFP_agp}
  T_{r,s}= (1-\alpha)^{s(s+1)/2} \sum_{0\leq m_1\leq\dots\leq
    m_s<r}\ \prod_{j=1}^s\alpha^{m_j} \ \prod_{1\leq j<k\leq s}
  \frac{2m_k-2m_j+k-j}{k-j}.
\end{equation}
This expression may remind the partition function of some discrete
log-gas, despite some relevant difference. Actually it exhibits
strong similarities with the partition function for pure continuous
Yang-Mills theory on the two-sphere \cite{DK-93}, and may similarly be
treated by resorting to random matrix model technologies.

\section{The six-vertex model and domino tilings}\label{sect.domino}

\begin{figure}
\centering
\includegraphics{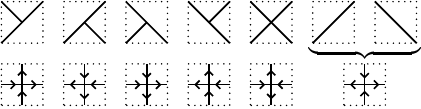}
\caption{Patches of domino tilings (first row) and 
corresponding vertices of the six-vertex model (second row).}
\label{fig-DominoPatches}
\end{figure}

\begin{figure}
\centering
\includegraphics{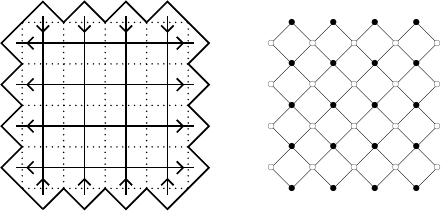}
\caption{An Aztec diamond of order $4$ and the underlying $4\times 4$
lattice with DWBC (left), and the related bipartite graph for 
dimer coverings (right).}
\label{fig-AztecDiamond}
\end{figure}

Let us here recall the correspondence between the six-vertex model on
a square lattice and the domino tilings \cite{EKLP-92a,EKLP-92b}. Given
a generic tiling, it may be decomposed into elementary patches, that
may be mapped into arrow configurations of the six-vertex model as
shown in \figurename~\ref{fig-DominoPatches}. The Aztec diamond of 
order $N$ then
corresponds to the six-vertex model on an $N\times N$ lattice with
DWBC, see \figurename~\ref{fig-AztecDiamond}.

From \figurename~\ref{fig-DominoPatches}, it is clear that two different
elementary patches of domino tilings correspond to the same one vertex
of type 6.  It follows that the enumeration of all possible tilings of
the Aztec diamond by dominoes is equal to the partition function of
the six-vertex model with DWBC, with the following choice of
Boltzmann weights: $w_1=\dots=w_5=1$, and $w_6=2$.  Indeed, with this
choice of the weights one readily recovers from \eqref{Z_FF} the celebrated
result $2^{N(N+1)/2}$, proven in \cite{EKLP-92a,EKLP-92b}.

Clearly, with the above choice, the Boltzmann weights satisfy the
free-fermion condition \eqref{FF}, and correspond to the
specialization $\rho=2$ and $\alpha=1/2$ in the parametrization
\eqref{weightsFF}.  As mentioned above, $\rho$ is simply  an overall
normalization of the Boltzmann weights, while $\alpha\in[0,1]$ is
relevant, parameterizing the different weights $\sqrt{2(1-\alpha)}$ or
$\sqrt{2\alpha}$, given to dominoes according to their NE-SW, or NW-SE
orientation, respectively \cite{JPS-98}.

\begin{figure}
\centering
\includegraphics{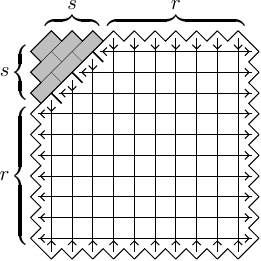}
\caption{The Aztec diamond of order $r+s$ of a pentagonal shape
and the corresponding lattice of the six-vertex model 
with a cut-off triangle, for $r=7$ and $s=3$. Note the comb-like structure on the
slanted side, constraining the position of dominoes along that
boundary, and induced in the domino tiling picture by the vertices of
type 2 in the triangular frozen region of the six-vertex model.}
\label{fig-CutoffTriangle}
\end{figure}

Given an Aztec diamond of order $N=r+s$, we may remove from it $s$
diagonal rows, $s<N$, starting from the top-left corner,
see \figurename~\ref{fig-CutoffTriangle}.  We shall refer to the
obtained region as the pentagonal domain, parameterized by the
integers $r$ and $s$, with $r=N-s$. Note that the cut-off triangular
domain admits only one tiling, with all dominoes oriented in the NE-SW
direction.

Consider now the six-vertex model on an $(r+s)\times (r+s)$ lattice
with a frozen triangular portion of size $s$ (consisting in $s(s+1)/2$
vertices of type 2) removed from the top-left corner.  By
construction, the resulting model still has DWBC, namely vertical
external arrows are all incoming, and horizontal ones are all
outgoing, including along the new boundary,
see \figurename~\ref{fig-CutoffTriangle}.  We shall denote by
$Z_{r,s}$ the partition function of the six-vertex model on such an
obtained lattice. Clearly, $Z_{r,0}=Z_r$.

Due to the relation between vertex and domino
configurations, \figurename~\ref{fig-DominoPatches}, the six-vertex
model on the restricted domain indeed translates into a pentagonal
domain, modulo a comb-like structure constraining the configurations
of dominoes adjacent to the slanted boundary,
see \figurename~\ref{fig-CutoffTriangle}. 

We want to study the limit of large lattice sizes, with large sizes of
the cut-off triangle, that is, we consider both $r$ and $s$ large,
with their ratio fixed.  For a later use, we introduce the parameter
\begin{equation}\label{omega}
\omega=\frac{s}{r+s}.
\end{equation}
Since $r$ and $s$ are positive integers, $\omega \in (0,1)$.  The free
energy per site of the model is defined as
\begin{equation}\label{free_tilings}
  F(\omega):=-\lim_{\substack{r,s\to\infty \\ s/(s+r)=\omega}}
  \frac{\log Z_{r,s}}{r^2+2rs+s(s-1)/2}.
\end{equation}
The partition function of the six-vertex model on the
pentagonal domain, $Z_{r,s}$, coincides, modulo a simple overall
factor, with the TDEFP, see \figurename~\ref{fig-GEFPbis}c. Indeed,
even under generic choice of the Boltzmann weights, it follows from
the definition of TDEFP that
\begin{equation}
  Z_{r,s}=\frac{Z_{r+s}}{w_2^{s(s+1)/2}}T_{r,s}.
\end{equation}
If we specialize the Boltzmann weights according to the
parameterization \eqref{weightsFF}, we get
\begin{equation}\label{z_efp}
Z_{r,s}=\frac{\rho^{(r+s)(r+s+1)/2-s(s+1)/4}}{(1-\alpha)^{s(s+1)/4}}T_{r,s}.
\end{equation}

In the scaling limit, that is, when $r$ and $s$ are both large, with
the ratio $\omega=s/(r+s)$ fixed, the leading behaviour of TDEFP is
described by the function
\begin{equation}\label{sigma}
\sigma(\omega):=-\lim_{s\to\infty}\frac{1}{s^2}\log T_{\lceil(1/\omega-1)s
\rceil,s}
\end{equation} 
Some properties of the function $\sigma(\omega)$ follow from the
definition of TDEFP. In particular, $T_{r,s}$ being a probability, it
varies between 0 and 1, implying $\sigma(\omega)\geq 0$. Moreover,
since the number of configurations contributing to the TDEFP may only
decrease (or stay constant) as the ratio $s/(r+s)$ increases,
$\sigma(\omega)$ is a non-decreasing function.

Recalling \eqref{Z_FF}, \eqref{weightsFF}, \eqref{z_efp}, and
\eqref{sigma}, we may thus write the free energy per site of the
six-vertex model on the pentagonal domain \eqref{free_tilings} as
follows
\begin{equation}
  F(\omega)=-\log\sqrt{\rho}+\frac{\omega^2}{4-2\omega^2}\log(1-\alpha)+ \frac{2\omega^2}{2-\omega^2}\sigma(\omega).
\end{equation}
The free-energy density $F(\omega)$ is therefore completely determined
by the function $\sigma(\omega)$, defined in \eqref{sigma}, whose
exact form  will be derived  below.

\begin{figure}
\centering
\includegraphics{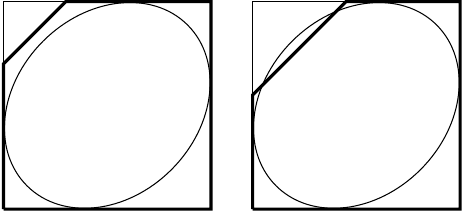}
\caption{Two scenarios for the thermodynamic limit, $r,s \to\infty$, shown schematically:  the triangular region lies outside of (left)
and overlaps with (right) the arctic ellipse of the original square domain. 
For the first case $\omega \in (0,\omegac]$ and for the second case 
$\omega\in[\omegac,1)$, where $\omega=r/(r+s)$ and $\omegac=1-\sqrt{\alpha}$. 
The border of the  triangular region exactly
touches  the ellipse at the critical value $\omega=\omegac$.}
\label{fig-ArcticEllipse}
\end{figure}

Let us conclude this Section by discussing the qualitative behaviour
of the function $\sigma(\omega)$.  Recall that the arctic circle (or
ellipse) phenomenon in the domino tilings of the Aztec diamond is the
appearance, in the scaling limit, of four regions of order, sharply
separated from a central region of disorder. In the language of the
six-vertex model, we thus have in the top-left corner, outside the
arctic ellipse, a region with all its vertices of type 2.  It follows,
from its very definition, that TDEFP must be close to one as long as
the triangular region of side $s$ entirely lies in the top-left
frozen region \cite{CP-07a}. In other words, as $r,s\to\infty$, the
TDEFP tends to $1$ when $r$ and $s$ are such that $s\ll r$, and tends
to $0$ when $r$ and $s$, being of the same magnitude, are such that
the triangular region overlaps with the central disordered region, 
see \figurename~\ref{fig-ArcticEllipse}.

Concerning the function $\sigma(\omega)$, defined in \eqref{sigma}, we
thus expect it to vanish identically for $\omega\in(0,\omegac]$, where
the critical value $\omegac=\omegac(\alpha)$ corresponds to the
diagonal side of the cut-off triangle being exactly tangent the arctic
ellipse.  From the known position of the arctic curve, corresponding
for TDEFP to $s\sim (1-\sqrt{\alpha})(r+s)$, we expect
\begin{equation}\label{omegac}
\omegac=1-\sqrt{\alpha}.
\end{equation}
On the remaining interval, $\omega\in [\omegac,1)$, we expect the
function $\sigma(\omega)$ to be positive-valued and non-decreasing. As
we shall see, these features will be indeed observed in the explicit  form
of the function $\sigma(\omega)$, whose exact derivation we shall now address.

\section{Scaling limit behaviour of TDEFP}

\subsection{Preliminaries}

Our starting point is representation \eqref{TDEFP_agp} for the TDEFP.
We have
\begin{equation}
  T_{r,s}=\left(\frac{1-\alpha}{\sqrt{\alpha}}\right)^{s(s+1)/2}
  g_{r,s},
\end{equation}
with
\begin{equation}
  g_{r,s}:=\alpha^{s(s+1)/4}\sum_{0\leq m_1\leq\dots\leq m_s<r}\ \prod_{j=1}^s
  \alpha^{m_j}\ \prod_{1\leq j<k\leq s} \frac{2m_k-2m_j+k-j}{k-j}.
\end{equation}
where we have extracted a simple $\alpha$-dependent factor for later
convenience.

It is convenient to introduce the `shifted' variable
\begin{equation}\label{shifted}
h_j:=2m_j+j, \qquad j=1,\dots, s,
\end{equation}
satisfying
\begin{equation}
  \frac{h_k-h_j}{k-j} \geq 1.
\end{equation}
in terms of which we may  write
\begin{equation}\label{tefp_rmm}
  g_{r,s}\propto \sum_{0< h_1<\dots <h_s<2r+s-1}\ \prod_{j=1}^s
  \alpha^{\frac{1}{2}h_j} \prod_{1\leq j<k\leq s} (h_k-h_j).
\end{equation}
Note that in the multiple sum, the $h_j $ runs over even (odd)
integers for even (odd) values of $j=1,\dots,s$. In \eqref{tefp_rmm}
we are ignoring an overall pre-factor, depending neither on $r$
nor on $\alpha$, but only on $s$, and whose explicit values is
irrelevant for our purpose.

Now, since $h_j$ is strictly increasing, the Vandermonde-like term is
always positive (or vanishing, when $h_k=h_j$), and we may as well
take its absolute value, obtaining a summand which is symmetric under
interchange of its variables. However, the symmetrization of the sum
is non-trivial, due to the constraints on the parity of its
indices. Ignoring these constraints, as they apparently may have only 
a sub-leading effect in addressing the
asymptotic behaviour of \eqref{tefp_rmm} in the scaling limit, we thus
can write
\begin{equation}\label{tefp_rmm2}
g_{r,s}\sim\kappa_s
\sum_{0< h_1,\dots, h_s<2r+s-1}\ \prod_{j=1}^s \alpha^{\frac{1}{2}h_j}
\prod_{1\leq j<k\leq s} |h_k-h_j| .
\end{equation}
Here, now the $h_j$s run over all the discrete interval $[1,2r+s-2]$
and $\kappa_s$ is some constant in $\alpha$, independent of 
$r$, but may depend on $s$.
 
Summing up, we have rewritten the TDEFP as the partition of a discrete
$\beta=1$ log-gas, consisting in $s$ particles on a lattice, at
positions $h_j$, $j=1,\dots,s$. The repulsive logarithmic repulsion
has a factor $1/2$ with respect to the usual $\beta=2$
case. The particles are confined to the interval $(0,2r+s-1)$ by two
`hard walls', and feel a linear potential
$-h\log\sqrt{\alpha}$. Recall that $\alpha\in(0,1)$, implying a
positive slope for the potential.

The large $s$ limit of $g_{r,s}$ can be described by the free energy
of the log-gas, $\Phi(\theta)$, defined by
\begin{equation}\label{Phi}
\Phi(\theta):=-\lim_{s\to\infty}\ \frac{1}{s^2}\log g_{\lceil(\theta-1)s/2\rceil,s},
\end{equation}
where we have introduced the parameter $\theta:=(2r+s)/s$ varying over
$(1,\infty)$.  The functions $\sigma(\omega)$, defined
in \eqref{sigma}, and $\Phi(\theta)$ are related by
\begin{equation}\label{sigmaphi}
\sigma(\omega)=\frac{1}{2}\log\frac{\sqrt{\alpha}}{1-\alpha}+\Phi(\theta),
\end{equation}
with $\theta=2/\omega-1$.

\subsection{Limiting cases as $\alpha\to 0$ and $\alpha\to 1$}

We shall need below, for the full determination of the function
$\Phi(\theta)$, the knowledge of its behaviour in the two limiting cases
$\alpha\to 0$ and $\alpha\to 1$.

In the first case, $\alpha\to  0$, the weights $w_3$ and $w_4$ vanish,
and there is only one configuration contributing to both the numerator
and the denominator of TDEFP. Therefore, $T_{r,s}|_{\alpha=0}=1$,
implying the behaviour
\begin{equation}\label{grsto0}
  g_{r,s} \sim\alpha^{s(s+1)/4},\qquad \alpha\to  0,
\end{equation}
and thus
\begin{equation}\label{phito0}
  \Phi(\theta)\sim-\frac{1}{4}\log\alpha,\qquad \alpha\to 0.
\end{equation}

The evaluation of the second limit, $\alpha\to 1$, is slightly more
elaborate. This case corresponds to vanishing Boltzmann weights for
the vertices of type 1 and 2. It follows from the definition of TDEFP,
requiring to have $s(s+1)/2$ vertices of type 2 in a triangular
region in the top-left corner, and from the fact that vertices of type
1 and 2 always come in pairs, that
\begin{equation}
  T_{r,s}\sim (1-\alpha)^{s(s+1)/2} C_{r,s}, \qquad \alpha\to 1, 
\end{equation}
where $C_{r,s}$ does not depend on $\alpha$, and thus
\begin{equation}
  \lim_{\alpha\to  1} g_{r,s}=C_{r,s}.
\end{equation}
Recalling now the determinant representation \eqref{TDEFP_det} for
$g_{r,s}$, we have:
\begin{align}
  C_{r,s}&=\det_{1\leq j,k \leq
    s}\left[\res_{x=1}\frac{x^{r+j+k-2}}{(x-1)^{2j}} \right]
  \\
  &=\det_{1\leq j,k \leq s}\left[
    \binom{r+j+k-2}{r-j+k-1}\right]
  \\
  &=(-1)^{s(s-1)/2} \det_{1\leq j,k \leq s}
  \left[ \binom{r+s+k-j-1}{r-s+k+j-2}\right].
\end{align}
The determinant may be  evaluated by resorting to \cite{Kr-01}, see
Theorem 27 therein, yielding:
\begin{equation}
  C_{r,s}=\binom{r+s-1}{s}\prod_{j=1}^s \frac{j!}{(2j-1)!}\
  \prod_{1\leq j <k\leq s}(2r+j+k-2).
\end{equation}

We are interested into the behaviour of $C_{r,s}$ in the scaling
limit, $r,s\to\infty$, with $2r/s+1=\theta$. Let us define
\begin{equation}
  \psi(\theta):=-\lim_{s\to\infty}\ \frac{1}{s^2}
  \log C_{\lceil(\theta-1)s/2\rceil,s}\,.
\end{equation}
Using standard  arguments based on Stirling formula, we obtain
\begin{equation}\label{psi}
  \psi(\theta):=-\frac{1}{4}\left[(\theta+1)^2
  \log(\theta+1)-2\theta^2\log\theta
    +(\theta-1)^2\log(\theta-1)\right]
  +\log 2.
\end{equation}
We thus have
\begin{equation}\label{phito1}
  \lim_{\alpha\to 1}\Phi(\theta)
  =\psi(\theta),
\end{equation}
with $\psi(\theta)$ given  by \eqref{psi}.

\subsection{Solution of the log-gas model}

To evaluate the function $\Phi(\theta)$, we follow \cite{DK-93} and
introduce the re-scaled variables
\begin{equation}
  \mu(x)=\frac{h_j}{s},\qquad x=\frac{j}{s}.
\end{equation}
These variables satisfy the condition
\begin{equation}\label{newconstraint}
\frac{\mu(y)-\mu(x)}{y-x}\geq 1.  
\end{equation}
Sums may now be interpreted as Riemann sums, which, in the large $s$
limit, turns into integrals. We thus have
\begin{equation}
  g_{r,s}\propto  \int \mathcal{D}[\mu(x)]\rme^{-s^2S[\mu(x)]},
  \qquad r,s\to\infty,
\end{equation}
with the action
\begin{equation}
S[\mu(x)]=-\frac{1}{2}\int_0^1\int_0^1
        \log\big\vert\mu(y)-\mu(x)\big\vert\rmd x\rmd y
        -\frac{1}{2}\log\alpha\int_0^1  \mu(x)\rmd x.
\end{equation}
Note that, compared to the situation treated in \cite{CP-13}, we have
a factor $1/2$ related to the $\beta=1$ nature of the considered
log-gas, and a factor $1/2$ associated to the change of
variable \eqref{shifted}. As a result, the action has an overall $1/2$
factor, and the corresponding saddle-point equation will therefore
exactly coincide with the $\beta=2$ case treated in \cite{CP-13}.

We now introduce the density
\begin{equation}
  \rho(\mu)=\frac{\partial x(\mu)}{\partial\mu}
\end{equation}
which must satisfy
\begin{equation}
\int_S\rho(\mu)\rmd\mu=1,\qquad \rho(\mu)\leq 1.
\end{equation}
where $S$ is the support of the density, $S\subset[0,\theta]$.  The
first equation is the usual normalization condition, while the second
one implements the constraint \eqref{newconstraint}.

Next, we introduce the resolvent, which, in the saddle-point
approximation, is related to the eigenvalue density as follows
\begin{equation}
W(z)=\int_S \frac{\rho(\mu)}{z-\mu}\rmd\mu,\qquad z\not \in S.
\end{equation}
Equivalently, we may express the density as the discontinuity of the
resolvent across its cut $S$,
\begin{equation}
  \rho(z)=-\frac{1}{2\pi\rmi}[W(z+\rmi 0)-W(z-\rmi 0)],\qquad z \in S.
\end{equation}
Clearly,  $S$ is also the support of the density.

In terms of the resolvent,  the saddle-point equation reads
\begin{equation}\label{SPE}
  W(z+\rmi 0)+W(z-\rmi 0)=2 U(z), \qquad z \in S,
\end{equation}
where, in absence of saturation, $U(z)$ is just the derivative of the
potential (i.e., simply $-\log\sqrt{\alpha}$, in our case). The factor
$2$ in the RHS come from the $\beta=1$ ensemble of our log-gas.  The
normalization condition for the density implies for the resolvent the
following asymptotic behaviour:
\begin{equation}\label{large_z}
  W(z)\sim\frac{1}{z}+\frac{E}{z^2}+ O(z^{-3}), \qquad\vert
  z\vert\to\infty,
\end{equation}
here $E$ is the the first moment of the density, or, equivalently, the
average of the eigenvalues.  It is worth emphasizing that, for a
linear potential, as it is the case here, the average $E$ and the
function $\Phi(\theta)$, see \eqref{Phi}, are simply related by
\begin{equation}\label{E-Phi}
-2\alpha\partial_\alpha\Phi(\theta)=E.
\end{equation}
Thus the knowledge of the next-to-leading order in the asymptotic
expansion of the resolvent at large $|z|$ allows one to determine
$\Phi(\theta)$ up to some $\alpha$-independent quantity.

The presently considered case, with exactly the same saddle-point
equation, has been treated in \cite{CP-13}.  Due
to the linear potential with positive slope, the particles tends to
accumulate to the left. As a consequence, we always observe saturation
in the vicinity of the origin. Two different scenarios may occur.
Given $a$, $b$, with $0<a<b<\theta$, we have:
\begin{itemize}
\item \emph{Scenario I} --- If
the right hard wall is sufficiently far away, there is a saturation for
$\mu\in[0,a]$, a band for $\mu\in[a,b]$, and a void for
$\mu\in[b,\theta]$;
\item 
\emph{Scenario II} --- If the right hard wall is
relatively closer, there are two saturated regions for
$\mu\in[0,a]\cup[b,\theta]$, and a band for $\mu\in[a,b]$.
\end{itemize}
For assigned values of $\alpha$, a phase transition occurs when
$\theta=b$. We shall verify below that this condition indeed reproduces
the critical value
\begin{equation}\label{thetac}
\thetac=\frac{2}{\omegac}-1=\frac{1+\sqrt{\alpha}}{1-\sqrt{\alpha}}
\end{equation}
expected from the correspondence  with domino tilings, see \eqref{omegac}.

Before proceeding we note that our saddle-point equation is exactly
the same as in the case of the square EFP \cite{CP-13}. Indeed, the
factor 2 in the RHS is compensated by the factor $1/2$ in the
potential.  Conversely, the modifications to the saddle-point equation
induced by the possible presence of saturated regions are not affected
by this factor 2. But the final form of $\Phi(\theta)$ is, due to the
different `boundary conditions' at $\alpha=0,1$, with respect to the
case of the square EFP.

\subsubsection{Scenario I} 

In this case the expression for the resolvent
reads
\begin{equation}\label{WI}
W_\mathrm{I}(z)=-\log\sqrt\alpha
-2\log\frac{\sqrt{a(z-b)}+\sqrt{b(z-a)}}{\sqrt{(b-a)z}}.
\end{equation}
The constant $a$ and  $b$ are determined by  imposing
the asymptotic behaviour \eqref{large_z} to the resolvent. We have
\begin{equation}
  a=\frac{1-\sqrt{\alpha}}{1+\sqrt{\alpha}}, \qquad
  b=\frac{1+\sqrt{\alpha}}{1-\sqrt{\alpha}}.
\end{equation}
This scenario requires the presence of a void in the interval
$[0,\theta]$. Thus, as $\theta$ is decreased, it is clear that the current
scenario breaks down at $\theta=\thetac\equiv b$, as anticipated
above.

Evaluation of the next order in the asymptotic behaviour of the resolvent
gives
\begin{equation}
E_{\mathrm{I}}=\frac{1+\alpha}{2(1-\alpha)},
\end{equation}
which, after integration, recall \eqref{E-Phi}, yields
\begin{equation}\label{phi_I}
  \Phi_{\mathrm{I}}(\theta)=-\frac{1}{2}\log\frac{\sqrt{\alpha}}{1-\alpha}.
\end{equation}
Note that the condition \eqref{phito0} fixes the value of the
integration constant to zero.

\subsubsection{Scenario II} In this case the expression for the resolvent
has the following form:
\begin{equation}\label{WII}
W_\mathrm{II}(z)=-\log\sqrt\alpha
-\log\frac{z-\theta}{z}
-2\log\frac{\sqrt{a(z-b)}+\sqrt{b(z-a)}}
{\sqrt{\theta-a}\sqrt{z-b}+\sqrt{\theta-b}\sqrt{z-a}}.
\end{equation}
The requirement of the asymptotic behavior \eqref{large_z} for the
resolvent leads the following set of equations for the end-points $a$ and $b$:
\begin{equation}
\begin{split}
\bigg(\frac{\sqrt{\theta-a}+\sqrt{\theta-b}}{\sqrt{b}+\sqrt{a}}\bigg)^2
&=\sqrt{\alpha},
\\
\sqrt{ab}+\sqrt{(\theta-a)(\theta-b)}
&=1.
\end{split}
\end{equation}
The solution reads
\begin{equation}\label{abII}
  a=\frac{\big(\sqrt{\theta+1}-
    \sqrt{(\theta-1)\sqrt{\alpha}}\big)^2}{2(1+\sqrt{\alpha})},
\qquad
b=\frac{\big(\sqrt{\theta+1}+
  \sqrt{(\theta-1)\sqrt{\alpha}}\big)^2}{2(1+\sqrt{\alpha})}.
\end{equation}
The condition $\theta= b$ determines the critical value $\thetac$,
 which indeed reproduces \eqref{thetac}.

Evaluation of the next order in the large $|z|$  behaviour of the
resolvent yields:
\begin{equation}
E_{\mathrm{II}}=
\frac{a+b}{4}+\frac{\theta}{2}\sqrt{(\theta-a)(\theta-b)}.
\end{equation}
Substituting here $a$ and $b$ from \eqref{abII}, and integrating,
see \eqref{E-Phi}, we obtain
\begin{equation}
  \Phi_{\mathrm{II}}(\theta)=
  \frac{1}{2}(\theta^2-1)\log\frac{2\alpha^{1/4}}{1+\sqrt{\alpha}}-
  \frac{1}{4}\theta\log\alpha +\psi(\theta),
\end{equation}
where $\psi(\theta)$ is given in \eqref{psi}.  Note that the
condition \eqref{phito1} fixes the value of the integration constant
exactly to $\psi(\theta)$. Expressing $\alpha$ in terms of $\thetac$, we
finally get
\begin{multline}\label{phi_II}
\Phi_{\mathrm{II}}(\theta)=
\frac{1}{2}\theta^2\log\frac{\theta}{\thetac}
-\frac{1}{4}(\theta-1)^2
\log\frac{\theta-1}{\thetac-1}
-\frac{1}{4}(\theta+1)^2
\log\frac{\theta+1}{\thetac+1}
\\
+\frac{1}{2}\log \frac{4\thetac}{\thetac^2-1}. 
\end{multline}
It may be verified that $\Phi(\theta)$ is continuous at
$\theta=\thetac$, together with its first and second derivatives, but
has a discontinuity in its third derivative. Such third-order
phase transition, in association to the discreteness of a log-gas model,
is known as Douglas-Kazakov phase transition \cite{DK-93}.

\section{Discussion}

Recalling that functions $\sigma(\omega)$ and $\phi(\theta)$ are related
by \eqref{sigmaphi}, our expressions \eqref{phi_I} and \eqref{phi_II}
implies:
\begin{equation}
\sigma(\omega)=0,\qquad  \omega\in(0,\omegac],
\end{equation}
and
\begin{equation}
\sigma(\omega)=
\frac{1}{2}\log\frac{\omega}{\omegac}
-\left(\frac{\omega-1}{\omega}\right)^2\log\frac{\omega-1}{\omegac-1}
+\frac{1}{2}\left(\frac{\omega-2}{\omega}\right)^2
\log\frac{\omega-2}{\omegac-2},
\qquad
\omega\in[\omegac,1).
\end{equation}
The function $\sigma(\omega)$ changes its behaviour at the critical
value $\omegac=1-\sqrt{\alpha}$, corresponding to geometric parameters
of the pentagonal domain such that the diagonal boundary of the frozen
triangle is exactly tangent to the arctic curve of the model.

The leading behaviour of $\sigma(\omega)$ in the vicinity of the point
$\omega=\omegac$, 
\begin{equation}
\sigma(\omega)\sim  \frac{1}{3(\omegac-2)(\omegac-1)\omegac^3}
(\omega-\omegac)^3, \qquad \omega\searrow\omegac,
\end{equation}
implies a discontinuity in the third-order derivative. Hence, the
function $F(\omega)$, defined in \eqref{free_tilings}, has also a
discontinuity in its third derivative at $\omega=\omegac$.  In other
words, as the scaled size of the cut-off triangle, $\omega$, is varied
across the critical value $\omegac$, a third-order phase transition
occurs in the model.

A similar result had already been derived in \cite{CP-13,CP-15}, for
an L-shaped domain. In that case the cut-off region was a square, or
rectangle, and the phase-transition was again triggered by the contact
between the cut-off region and the arctic curve.  The present result
provides further support to the interpretation of the phase separation
curves as critical curves in the space of parameters describing the
macroscopic geometry of the tiled domain.

The two above mentioned examples of third-order phase transitions
observed in domino tilings are clear consequences of the tight
connection of these model with discrete log-gases. The mechanism
underlying these phase transitions is closely related to those
unveiled in \cite{GW-80,W-80,DK-93} for random matrix models, that
have actually reappeared in many different models. See,
e.g., \cite{MS-14} for a review.

Concerning the pentagonal domain studied here, note that it slightly
differs from that considered in \cite{FV-21}. Indeed, in the present
case, the boundary conditions inherited from the six-vertex model
introduce some additional comb-like structure along the slanted side,
which slightly restrict the possible tilings of the domain,
see \figurename~\ref{fig-CutoffTriangle}. Denoting by
$\tilde{Z}_{r,s}$ the partition function for the random tilings of the
pentagonal domain with no such restriction, one can easily see verify
that $\tilde{Z}_{r-1,s+1} <Z_{r,s}<\tilde{Z}_{r,s}$. The microscopic
difference between the two domains becomes therefore negligible in
the scaling limit considered here.

As a last remark, note that the log-gas representation for EFP,
see \cite{CP-13}, Eq.~(3.4), is in the $\beta=2$ ensemble, while the
representation \eqref{TDEFP_agp} for TDEFP is in the $\beta=1$
ensemble. This is consistent with the fact that EFP tests the
fluctuations of the \emph{intersection} of the arctic curve with the
main diagonal, while TDEFP tests the fluctuations of
the \emph{maximum} of the arctic curve in the vicinity of the main
diagonal.  Indeed, these two statistics are known to be governed by
the Airy-2 and Airy-1 processes,
respectively \cite{J-05,CQR-13,FV-21}.

\section*{Acknowledgments} 

We are indebted to P. Di Francesco and
G. Panova for stimulating discussions. Part of this research was
performed while FC was visiting the Institute for Pure and Applied
Mathematics (IPAM), which is supported by the National Science
Foundation (Grant No. DMS-1925919).  
AGP acknowledges support from the Theoretical Physics and
Mathematics Advancement Foundation BASIS and from the Euler International Mathematical Institute grant No. 075-15-2022-289.

\bibliography{tdefp_bib}
\end{document}